# Satellite capture mechanism in a sun-planet-binary four-body system

Shengping Gong[*], Miao Li [†]

School of Aerospace Engineering, Tsinghua University, Beijing China, 100084

**Abstract**

This paper studies the binary disruption problem and asteroid capture mechanism in a sun-planet-binary four-body system. Firstly, the binary disruption condition is studied and the result shows that the binary is always disrupted at the perigee of their orbit instantaneously. Secondly, an analytic expression to describe the energy exchange between the binary is derived based on the 'instantaneous disruption' hypothesis. The analytic result is validated through numerical integration. We obtain the energy exchange in encounters simultaneously by the analytic expression and numerical integration. The maximum deviation of this two results is always less than 25% and the mean deviation is about 8.69%. The analytic expression can give us an intuitive description of the energy exchange between the binary. It indicates that the energy change depends on the hyperbolic shape of the binary orbit with respect to the planet, the masses of planet and the primary member of the binary, the binary phase at perigee. We can illustrate the capture/escape processes and give the capture/escape region of the binary clearly by numerical simulation. We analyse the influence of some critical

---

[*] Associate professor, School of Aerospace Engineering; gongsp@tsinghua.edu.cn

† PhD candidate, School of Aerospace Engineering; 15652773694@163.com

factors to the capture region finally.

**Key words**: binary; capture mechanism; disruption; analytic derivation; numerical results

# 1 Introduction

In the last decade, hundreds of irregular satellites orbiting giant planets have been found in our solar system. Irregular satellites characterized by large orbits, with high eccentricity and inclinations are in the majority. The origin of these irregular satellites is a controversial and popular topic, notably Neptune's moon (Agnor and Hamilton 2006), and Jupiter's moon (Nesvorny et al. 2007; Philpotta et al. 2010). Besides, the old idea of capturing an asteroid to ease the Earth's natural resources pressure becomes realizable as the deepening study (Hasnain et al. 2012; Mazanek et al. 2013). The motivations of capturing an asteroid include exploiting its abundant natural resources and supporting the researches aimed at exploring the origin of the solar system (McAndrews et al. 2003). There is a popular belief that irregular satellites were captured by their planets, but detailed capture mechanisms are still undefined (Phipotta et al. 2010). Some popular models have been proposed, such as energy dissipation, pull-down, three-body interactions and collision. However, none of these models is a flawless theory to explain all scenarios.

Some particular physical processes are the main cause of the energy variation of satellites in capture. Three 'classical' mechanisms to explain the energy dissipation in capture have been proposed. First, capture by gas drag (Pollack et al. 1979), where the satellites are decelerated while passing through the gas disk and circumplanetary nebula surrounding a planet. However, for this mechanism to be efficient, the gas must be sufficiently dense and the asteroid should have encountered approximately its own mass within the nebula (Jewitt et al. 2005). This model is possible but is only suitable for very small satellites. Second, pull-down capture (Heppenheimer et al. 1977), where satellites are captured because of rapid enlargement of the planet's Hill sphere in a short time, satellites are captured when the planet's mass increases or the sun's mass decreases. The solar/planetary mass must change substantially and instantaneously to make this happen. However, the sun's mass decrease or planet growth require far more time than allowed by this capture scenario (Sheppard et al. 2005). Thus, in view of the law of planetary formation, this model is improbable and unpractical. Another mechanism involves collision capture (Colombo et al. 1971), which occurs when an asteroid collides with a planetary satellite or another asteroid in the vicinity of the planet. Cline (1979) studied the utilization of an existing satellite to capture an asteroid into a closed orbit about the planet. Due to the dynamic constraints, it has been proved that capture of retrograde satellites is virtually quite impossible in restricted three-body system (Tanikawa 1982). However, the asteroid–satellite scattering allows the asteroid to change both its angular momentum and energy with respect to the planet (Tsui 2000). The orbit distribution of the satellites captured by this mechanism provides a good match to the observations at Jupiter (Nesvorny et al. 2014). For this mechanism to be efficient, there must be sufficient passing bodies near the planets, far beyond the number presently observed but which might have occurred in the early solar system (Gomes et al. 2005; Hahn et al. 2005). Based on this three-body interaction model, some innovative theories and efficient models have been proposed, for example, an approaching binary system that encounters a planet. Some valuable work has been performed in recent years. A new theory of the origin of Jupiter's irregular satellite is studied. One member of a binary of about 100km is captured after tidal disruption (Phipotta et al. 2010). Binary planets (a binary system consisting of a pair of planets) can be formed by capture due to the planet–planet dynamic tide during orbital crossing of three giant planets (Ochiai et al. 2014). Planetary capture and escape after the satellite flyby in the planar four-body problem is researched (Gong et al. 2015). And several numerical examples are given to illustrate the satellite-aided capture.

Given that newly discovered abundance of binaries in small-body populations in the last decade (Stephens et al. 2006), the binary-planet encounters model is increasingly being accepted and is a likely mechanism for satellite capture (Morbidelli 2006). Satellite capture is dynamically possible in Sun-planet-binary four-body system even if the relative energy of the incoming binary with respect to the planet is large (Tsui 2001). The possibility of the binary-planet encounter mechanism was investigated by computational analysis (Maddison 2006). More than eight thousand numerical simulations of the Sun-Jupiter-binary four-body system have been performed to research the permanent capture mechanism (Gaspar et al. 2011). The capture probability by binary exchange relies on the statistics of the encounter and the relative energy (Nogueira et al. 2011), and the energy exchange in a simple binary disruption model can be evaluated (Kobayashi et al. 2012). One possible outcome of gravitational encounters between a binary-planet system might be an exchange reaction. The capture probability of the minor member of the binary is much greater than that of the major member (Gaspar et al. 2011). The distribution of the encounters two-planet system that is migrating due to interactions with an exterior planetesimal belt was numerically integrated (Quillen et al. 2012). They focused on the different performances between the inner planet and outer planet on the probability that a satellite is captured, and the probability of the irregular satellites are captured due to binary exchange is predicted, and the planetesimal binaries similar to those in the Kuiper belt would have a probability of 1/100 captured by an outer migrating planet. Agnor and Hamilton (2006) examined the capture of Triton by an exchange process between a binary pair and Neptune. In their theory, a binary was tidally disrupted and one of its members, Triton, was captured as a satellite. Numerical simulations for this particular case were used to illustrate the feasibility.

The main objective of this paper is to study the asteroid capture mechanism in the Sun-planet-binary system. Shiho Kobayashi et al. (2012) examined capture mechanism between the binary and the black hole at the Galactic center. The author discussed how the escape and capture preference between unequal-mass binary members depends on which orbits they approach the black hole. In view of the orbit of an asteroid around the planet is parabolic or elliptic before the encounter, the asteroid will be captured naturally. We only study the capture mechanism when a binary approaches the planet in hyperbolic orbit. In contrast to Agnor and Hamilton's capture model (2006), the analytic expression of energy exchange between the binary is obtained. The result of the analytic expression is compared to the numerical result, and the relative error is always less than 20%. The disruption/capture regions are investigated as well.

## 2  Planar four-body problem

The planar four-body problem, including Sun-planet-satellite-asteroid and Sun-planet-binary system, were investigated by Gong (2015) and Tsui (2001), respectively. The dynamical model in these references are used to study the binary disruption and capture problem. The four-body system includes the sun *S*, the planet *P*, and two asteroids *A* and *B*, where *A* and *B* form a binary system (see Figure 1). The planet orbits around the sun in a circular orbit. The motion of the binary system is studied when it approaches the planet. In the inertial frame of the Sun, the dynamical equations of *P*, *B* and *A*, are given by

$$\frac{\mathrm{d}^2 \mathbf{r}_A}{\mathrm{d}t^2} = -\frac{GM_S}{r_A^3}\mathbf{r}_A - \frac{GM_P}{r_{PA}^3}\mathbf{r}_{PA} - \frac{GM_B}{r_{BA}^3}\mathbf{r}_{BA} \quad (1)$$

$$\frac{\mathrm{d}^2 \mathbf{r}_B}{\mathrm{d}t^2} = -\frac{GM_S}{r_B^3}\mathbf{r}_M - \frac{GM_P}{r_{PB}^3}\mathbf{r}_{PB} + \frac{GM_A}{r_{BA}^3}\mathbf{r}_{BA} \quad (2)$$

$$\frac{\mathrm{d}^2 \mathbf{r}_P}{\mathrm{d}t^2} = -\frac{GM_S}{r_P^3}\mathbf{r}_P + \frac{GM_A}{r_{PA}^3}\mathbf{r}_{PA} + \frac{GM_B}{r_{PB}^3}\mathbf{r}_{PB} \quad (3)$$

Where *G* is the gravitational constant; $M_S$, $M_P$, $M_A$, and $M_B$ are the masses of the sun, planet, and the binary, respectively. We order $M_P = M_{Earth}$ in numerical simulation in this paper.

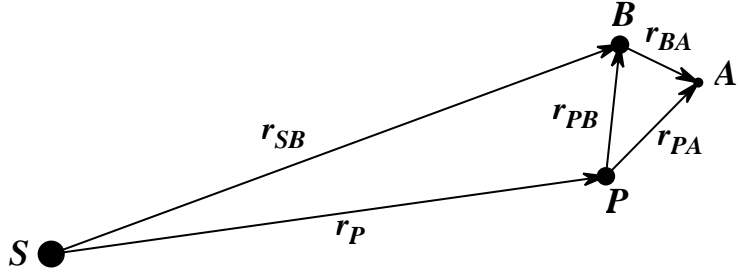

Figure 1. Geometry of the four-body system

Assume that the gravity of the sun dominates the system, the binaries do not influence the orbit of the planet and the distance between the asteroid and planet is small compared to the distance between the planet and the sun. Thus, the planet evolves on a Keplerian orbit around the sun. Therefore, the equation (3) can be simplified as

$$\frac{d^2 \bm{r}_P}{dt^2} = -\frac{GM_S}{r_P^3} \bm{r}_P \qquad (4)$$

The simultaneous equations of equations (1) and (4) give the dynamical equations of motion of the asteroid $A$ with respect to the planet.

$$\begin{aligned}\frac{d^2 \bm{r}_{PA}}{dt^2} &= \frac{d^2 \bm{r}_A}{dt^2} - \frac{d^2 \bm{r}_P}{dt^2} \\ &= -\frac{GM_S}{r_A^3} \bm{r}_A - \frac{GM_P}{r_{PA}^3} \bm{r}_{PA} - \frac{GM_B}{r_{BA}^3} \bm{r}_{BA} + \frac{GM_S}{r_P^3} \bm{r}_P\end{aligned} \qquad (5)$$

Because of $\bm{r}_{PA} \ll \bm{r}_P$, Then the right-hand side of equation (5) above can be linearized as

$$\frac{d^2 \bm{r}_{PA}}{dt^2} = -\left( \frac{GM_S}{r_P^3} \bm{I} - 3 \frac{GM_S \bm{r}_P \bm{r}_P^T}{r_P^5} \right) \bm{r}_{PA} - \frac{GM_P}{r_{PA}^3} \bm{r}_{PA} - \frac{GM_B}{r_{BA}^3} \bm{r}_{BA} \qquad (6)$$

Where $\bm{I}$ is an identity matrix.

The equation can be transformed to the planet-centred rotating frame. Assume that the planet rotates around the sun in a circular orbit. Then, the dynamical equation can

be written as

$$\ddot{\boldsymbol{r}}_{PA} + 2\boldsymbol{\omega}_P \times \dot{\boldsymbol{r}}_{PA} + \boldsymbol{\omega}_P \times (\boldsymbol{\omega}_P \times \boldsymbol{r}_{PA}) = \\ -\omega_P^2 \left(I - 3\frac{\boldsymbol{r}_P \boldsymbol{r}_P^T}{r_P^2}\right)\boldsymbol{r}_{PA} - \frac{GM_P}{r_{PA}^3}\boldsymbol{r}_{PA} - \frac{GM_B}{r_{BA}^3}\boldsymbol{r}_{BA} \quad (7)$$

Where $\boldsymbol{\omega}_P$ is the angular velocity of the planet around the sun.

The equation can be transformed into a planet-centred rotating frame. The origin of the frame is the mass centre of the planet; the *x*-axis points from the sun to the planet; the *z*-axis is along the direction of the angular momentum of the mutual rotation; and the *y*-axis forms a right triad with the *x* and *z* axes. Assume that the planet rotates around the sun in a circular orbit. The equation of motion allows for the nondimensionalization of the model and elimination of all free parameters. By taking the orbital radius of the planet around the sun $r_P$ as the unit length and $\tau = 1/\omega_P$ as the unit time, where $\omega_P$ is the angular velocity of the planet around the sun, the equation of motion of the asteroid *A* can be transformed into the following parameterless equations.

$$\begin{cases} \ddot{x}_A - 2\dot{y}_A = 3x_A - \dfrac{\mu_P}{r_{PA}^3}x_A - \dfrac{\mu_B}{r_{BA}^3}x_{BA} \\ \ddot{y}_A + 2\dot{x}_A = -\dfrac{\mu_P}{r_{PA}^3}y_A - \dfrac{\mu_B}{r_{BA}^3}y_{BA} \end{cases} \quad (8)$$

where $\mu_P = M_P/M_S$ is the dimensionless mass of the planet, and $\mu_B = M_B/M_S$ is the dimensionless mass of asteroid *B*.

Similarly, the equation of motion of asteroid *B* in the rotating frame can be given by

$$\begin{cases} \ddot{x}_B - 2\dot{y}_B = 3x_B - \dfrac{\mu_P}{r_{PB}^3}x_B + \dfrac{\mu_A}{r_{BA}^3}x_{BA} \\ \ddot{y}_B + 2\dot{x}_B = -\dfrac{\mu_P}{r_{PB}^3}y_B + \dfrac{\mu_A}{r_{BA}^3}y_{BA} \end{cases} \qquad (9)$$

where $\mu_A = M_A / M_S$ is the dimensionless mass of asteroid A.

Multiply the first equation in equation (8) by $\dot{x}_A$ and the second equation by $\dot{y}_A$. Then, adding them together gives

$$\dot{C}_A = -\dfrac{\mu_B}{r_{BA}^3}(\dot{x}_A x_{BA} + \dot{y}_A y_{BA}) \qquad (10)$$

where $C_A$ is the energy integral in the restricted three-body problem

$$C_A = \dfrac{1}{2}(\dot{x}_A^2 + \dot{y}_A^2) - \dfrac{3}{2}x_A^2 - \dfrac{\mu_P}{r_{PA}} \qquad (11)$$

Similarly, we can obtain the energy integral for asteroid B as

$$\dot{C}_B = \dfrac{\mu_A}{r_{BA}^3}(\dot{x}_B x_{BA} + \dot{y}_B y_{BA}) \qquad (12)$$

where

$$C_B = \dfrac{1}{2}(\dot{x}_B^2 + \dot{y}_B^2) - \dfrac{3}{2}x_B^2 - \dfrac{\mu_P}{r_{PB}}$$

Considering a standard sun–planet–asteroid system by setting $\mu_B = 0$ in equation (8), the system degenerates to a restricted three-body system, and the energy integral is constant. In this case, we can obtain the positions of two collinear Lagrange points, $x_{1,2} = \pm(\mu_P/3)^{1/3}$. Letting the velocity be zero, equation (11) defines the zero velocity curves, which are the bounds of the motion of the asteroid. The energy constant at the collinear Lagrange point $L_1$ is denoted as $C_1$. The motion is restricted around the sun or planet, depending on its initial states, if $C < C_1$. When asteroid A is out of the SOI (Sphere of Influence) of asteroid B, the sun-planet-binary system degenerates to a sun-planet–asteroid system. Therefore, asteroid A is captured by the planet if the energy integral is smaller than $C_1$ without the gravitational force of satellite B.

Disruption of the binary is the precondition of this capture model. The results in Gaspar (2010) show that the permanent capture probability of the minor member of the binary is greater than the major body permanent capture probability. This paper focuses on the capture of the minor member of the binary. Assume the mass of asteroid A is far less than the mass of asteroid B, namely, $M_A \ll M_B$. Thus, asteroid A is assumed to rotate around B in a circular orbit. Because of $M_A \ll M_B \ll M_P$, the binary is disrupted if asteroid A is out of the Hill sphere of B after encountering. Further, if asteroid A orbits around the planet stably after the encounter, it has been captured by the planet.

## 3 Disruption of the binary

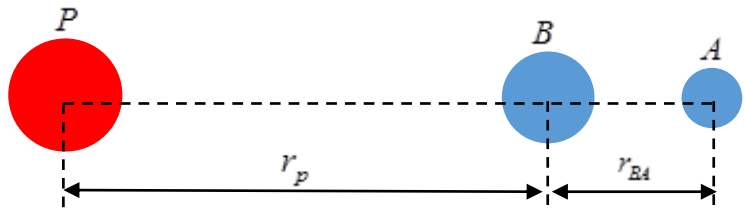

Figure 2. Schematic diagram of the binary disruption

The binary orbits around the planet in a hyperbolic trajectory in the SOI (sphere of influence) of the planet. Based on the impulse approximation proposed by Agnor and Hamilton (2006), the disruption is instantaneous. The binary is disrupted if asteroid *A* is out of the Hill sphere of *B*, from which we can estimate the tidal disruption radius as

$$r_{td} \approx r_{AB}\left(\frac{3M_P}{M_A+M_B}\right)^{1/3} \approx r_{AB}\left(\frac{3M_P}{M_B}\right)^{1/3} \tag{13}$$

where $r_{AB}$ is the separation between the asteroids.

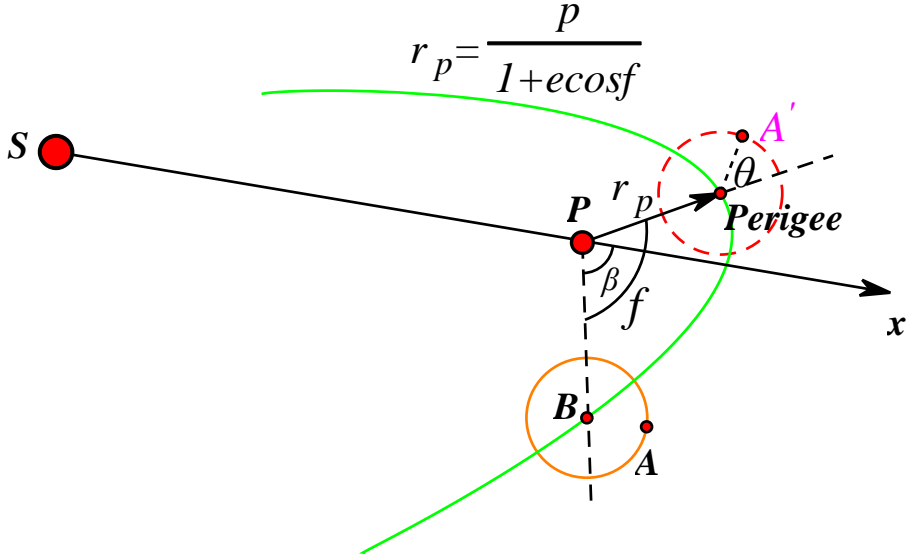

Figure 3. Schematic diagram when the binary is in the SOI of the planet

Figure 3 illustrates the orbits of the asteroids in the vicinity of the planet, where $\theta$ is the phase angle between the asteroids, and $\theta_t$ is the angle at the perigee, $\beta$ is an angular variable which measured from the **PB** line to the **SP** line, and *f* is the true anomaly of the hyperbolic orbit, *e* is the eccentricity of the hyperbolic.

Equations (8) and (9) are integrated from the time when the binary enters the SOI of the planet to the time when they leave the SOI. The distance between the asteroids when they leave the SOI of the planet is used to judge the disruption. Numerical results for some particular cases are given to examine the conditions of binary disruption. If the distance is larger than the Hill radius, the binary breaks.

Two examples are given to illustrate the disruption and non-disruption cases for $\mu_P = \mu_{Earth}, \mu_B = 3\times 10^{-11}, \theta_t = 0.7\pi$. The other parameters are $e = 1.06$, $r_p = 0.8 r_{td}$, $r_{BA} = 2.3333 \times 10^{-6}$ AU and $e = 1.04$, $r_p = 0.9 r_{td}$, $r_{BA} = 2 \times 10^{-6}$ AU respectively. The separation between the asteroids and the trajectories of the asteroids departing from the planetary SOI are shown in Figure 4 and Figure 5.

As shown in Figure 4, the distance between asteroid $A$ and $B$ increases sharply when the binary reaches the perigee. Thus, the binary is disrupted in the vicinity of the perigee. In contrast, as shown in Figure 5, $r_{BA}$ is always less than the Hill sphere radius of asteroid $B$. This binary survives.

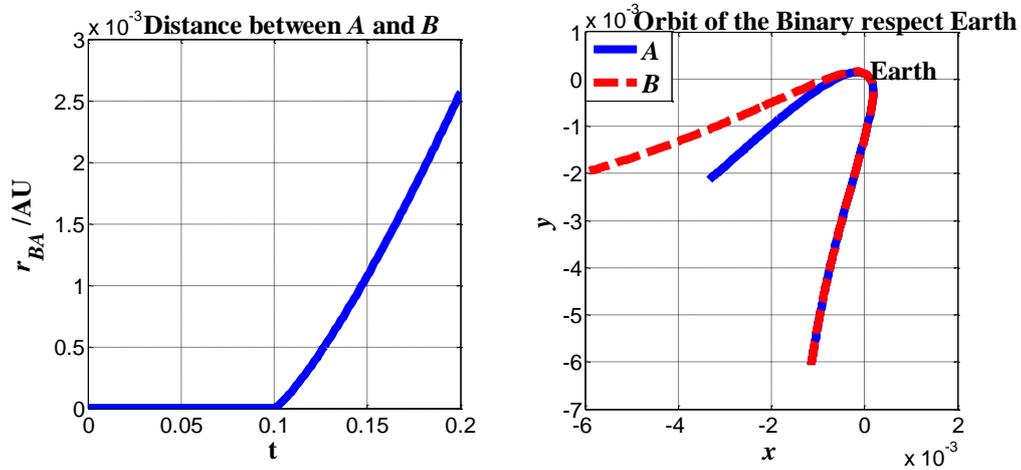

Figure 4. Disruption case for $r_p = 0.8 r_{td}$, $\mu_B = 3\times 10^{-11}$, $\theta_t = 0.7\pi$, $e = 1.06$, $r_{BA} = 2.3333\times 10^{-6}$

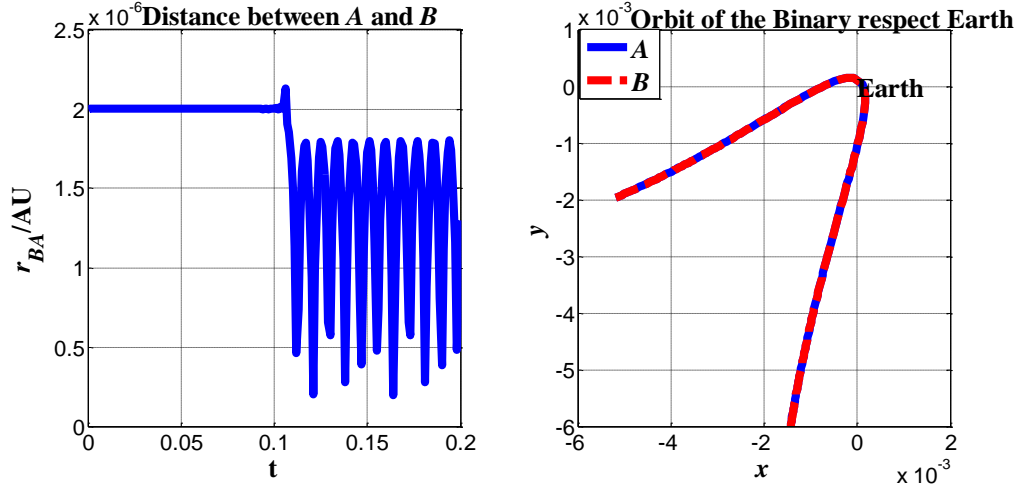

Figure 5. Unbroken case for $r_p = 0.9 r_{td}$, $\mu_B = 3 \times 10^{-11}$, $\theta_t = 0.7\pi$, $e = 1.04$, $r_{BA} = 2 \times 10^{-6}$

By logging lots of simulations of disruption, two assumptions have been verified:

(1) In this binary-planet encounter model, the binary disruption occurs instantaneously in the vicinity of the perigee.

(2) The distance $r_{BA}$ can be regarded as constant before the binary reaches the perigee.

**3.1 Disruption region**

In the instantaneous disruption model, the disruption is only dependent on the perigee radius. There are other parameters that influence the disruption condition, including the mass of the asteroid *B*, eccentricity *e* of the hyperbola, and phase angle between asteroids at the perigee. In this section, the disruption region is represented in the $e - r_p$ space at different values of $r_p$ and $\theta_t$.

1) **Influence of perigee radius $r_p$**

Based on the tidal disruption theory, the binary will be disrupted when $r_p < r_{td}$ is satisfied. However, disruption does not always occur as predicted, so it is necessary to study the influence of the perigee radius $r_p$ on binary disruption.

Given $\theta_t = 0.7\pi$, $\mu_B = 3\times 10^{-11}$, we can obtain the disruption regions for different values of $r_p = 0.6r_{td}$, $0.7r_{td}$, $0.8r_{td}$, $0.9r_{td}$, respectively, as shown in Figure 6. The disruption region expands when the perigee radius $r_p$ decreases and the probability of disruption approaches 100% when $r_p \leq 0.6r_{td}$.

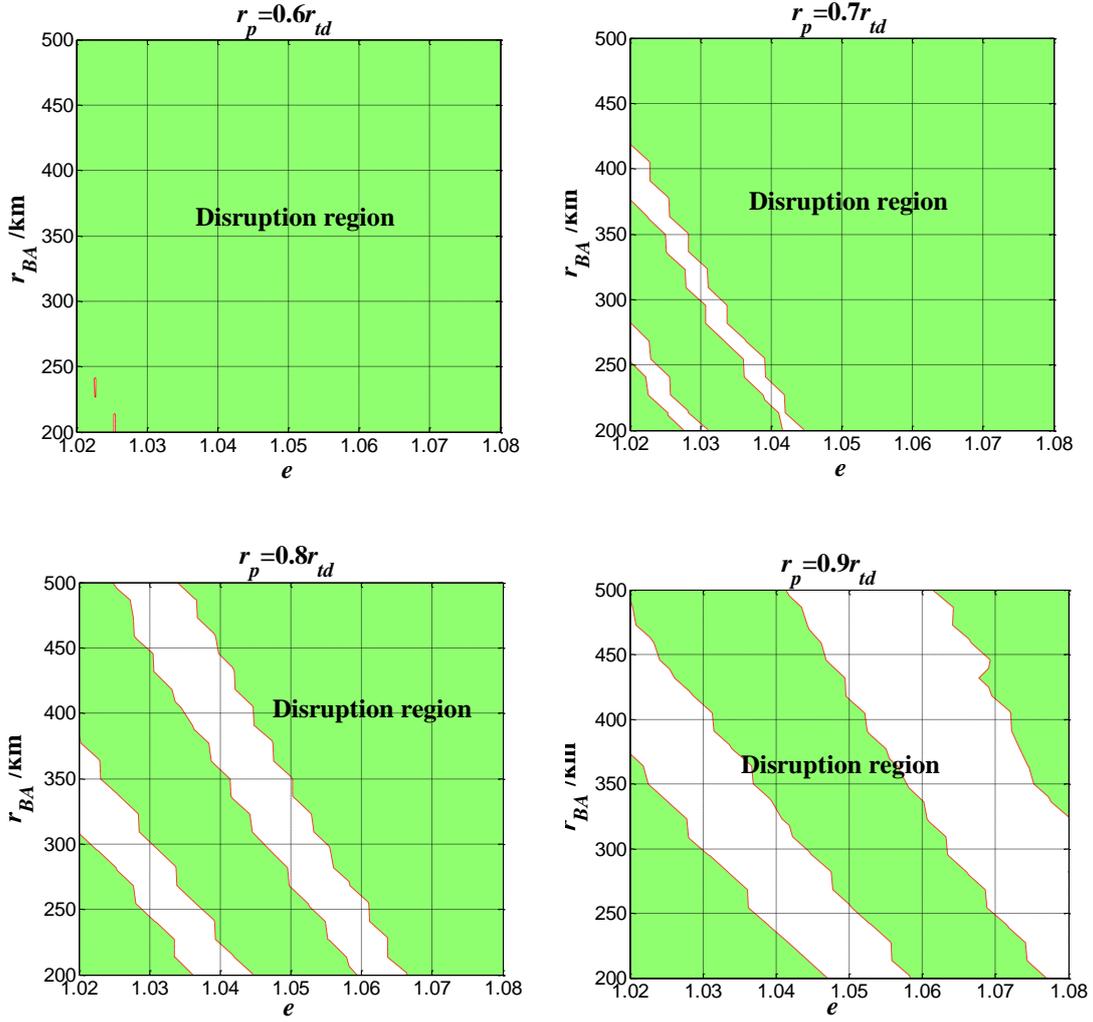

Figure 6. Disruption region for different values of perigee

## 2) Influence of phase angle $\theta_t$

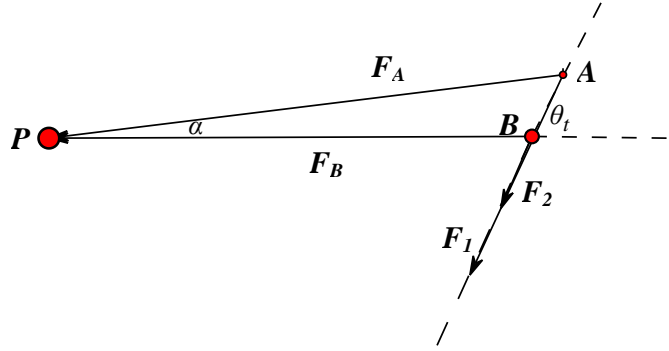

Figure 7. Mechanical model of the tidal disruption

As shown Figure 7, the tidal force at the perigee of the planet is given by:

$$F_{tid} = F_B \cos\theta_t - F_A \cos(\theta_t - \alpha) = F_1 - F_2 \qquad (14)$$

In view of $r_{BA} \ll r_{PB}(r_{PA})$, we can assume that $\alpha \approx 0$. Then, equation (14) can be simplified as

$$F_{tid} = (F_B - F_A)\cos\theta_t \qquad (15)$$

Theoretically, the tidal force $F_{tid}$ increases when $\theta_t$ approaches 0 or $\pi$. Thus, the disruption region expands when $\cos\theta_t$ approaches 1.

Given $r_p = 0.8 r_{td}, \mu_B = 3\times 10^{-11}$, disruption regions are obtained for different values of $\theta_t = 0.5\pi,\ 0.7\pi,\ 0.8\pi,\ 0.9\pi$, as shown in Figure 8. The disruption region expands when $\theta_t$ approaches $\pi$, as expected.

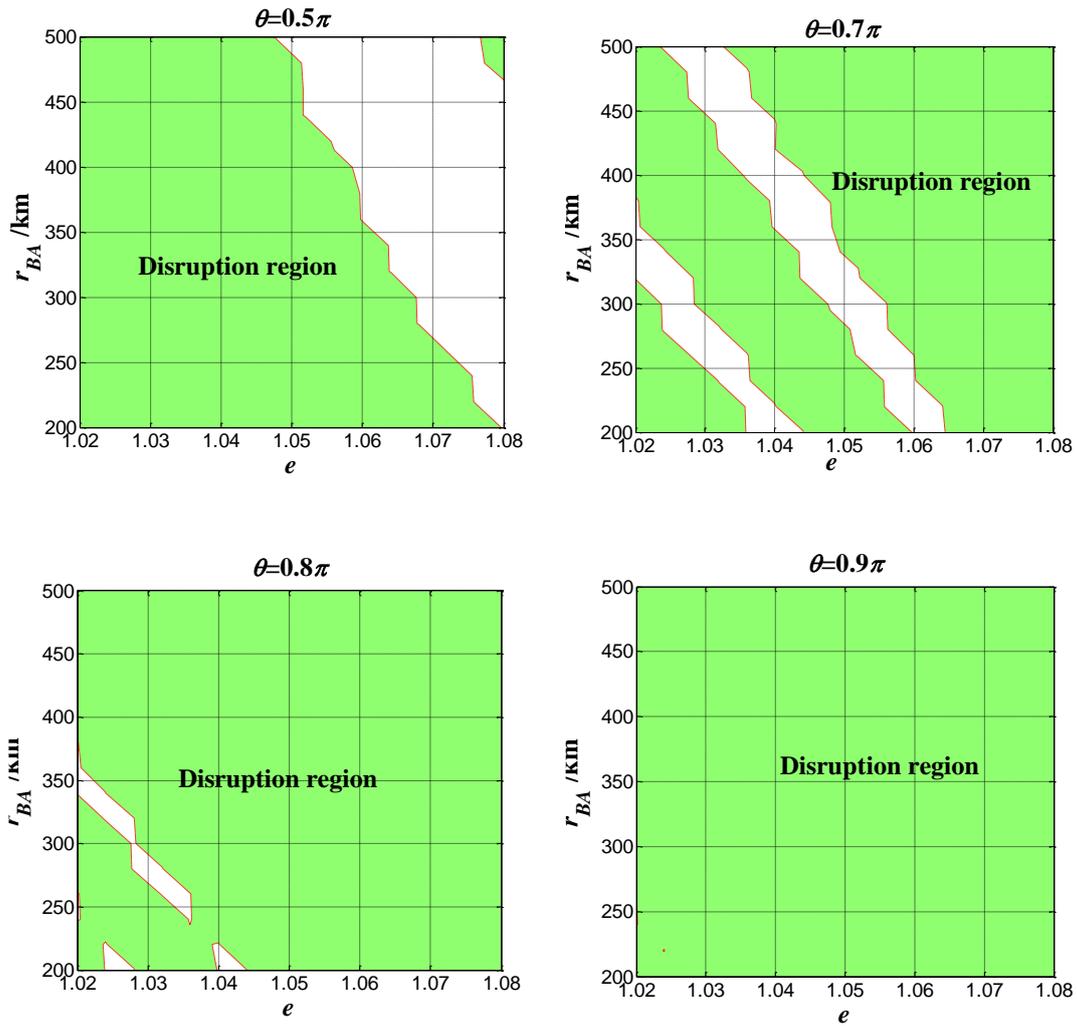

Figure 8. Disruption region of different phase $\theta_t$

## 4 Capture mechanism

Although the total energy of the sun-planet-binary system is conserved, the energy integral of the asteroids can be altered. Due to the interactions between the asteroids in equations (8) and (9), the energy exchange between asteroids *A* and *B* can result in the capture of one member of the binary. In principle, the smaller member of the binary has more chance to be captured for any hyperbolic orbit, because it carries a smaller fraction of the orbital energy (Kobayashi et al. 2012). During the tidal disruption, the probability that the smaller member of the binary is captured depends on the incoming velocity of the binary (Agnor and Hamilton 2006). The capture of the massive member is possible but the probability is low at higher velocity ($0.35 \leq v_\infty \leq 1.55 km/s$). The secondary member is more likely to be captured with low incoming velocity. Therefore, we mainly consider situation that the secondary member of the binary is captured.

We evaluate the energy variation analytically based on the assumption that the disruption is instantaneous and happens at the perigee. We need to simplify the expressions of $\dot{x}_A$, $\dot{y}_A$, $x_{BA}$, $y_{BA}$ in equation (10).

**4.1 Analytic derivation**

As shown in Figure 3, the binary flies by the planet in a hyperbolic orbit, and a two-body model is used to analyse the motion of the asteroid. To describe the position of asteroid $A$ with respect to asteroid $B$, a local reference frame $o_2 x_2 y_2 z_2$ is defined. The origin of the frame is the mass centre of the binary, where the $x_2$-axis is defined by the direction of the vector $\boldsymbol{PB}$, the $z_2$-axis is along the angular momentum direction of their mutual motion, and the $y_2$-axis forms a right triad with the $x_2$ and $z_2$ axes.

$$\begin{bmatrix} x_B \\ y_B \end{bmatrix} = r_{PB} \begin{bmatrix} \cos(\beta + f) \\ \sin(\beta + f) \end{bmatrix} \tag{16}$$

The $x_B, y_B$ in equation (16) express the position of $B$ respect to planet in the planet-centred rotating frame.

$$\begin{cases} \tilde{x}_{BA} = r_{BA} \cos\theta \\ \tilde{y}_{BA} = r_{BA} \sin\theta \end{cases} \tag{17}$$

$\tilde{x}_{BA}, \tilde{y}_{BA}$ in equation (17) express the position of $A$ respect to $B$ in the local reference frame.

The position can be expressed in the rotating frame by a coordinate transformation.

$$\begin{bmatrix} x_{BA} \\ y_{BA} \end{bmatrix} = \begin{bmatrix} \cos(\beta+f) & -\sin(\beta+f) \\ \sin(\beta+f) & \cos(\beta+f) \end{bmatrix} \begin{bmatrix} \tilde{x}_{BA} \\ \tilde{y}_{BA} \end{bmatrix} = r_{BA} \begin{bmatrix} \cos(\beta+\theta+f) \\ \sin(\beta+\theta+f) \end{bmatrix} \quad (18)$$

$$\begin{bmatrix} x_A \\ y_A \end{bmatrix} = r_{PB} \begin{bmatrix} \cos(\beta+f) \\ \sin(\beta+f) \end{bmatrix} + r_{BA} \begin{bmatrix} \cos(\beta+\theta+f) \\ \sin(\beta+\theta+f) \end{bmatrix} \quad (19)$$

The velocity of asteroid A with respect to the planet can be expressed by the summation of the velocity of asteroid A with respect to asteroid B and the velocity of asteroid B with respect to the planet.

$$\frac{d\boldsymbol{r}_{PA}}{d\tau} = \frac{d\boldsymbol{r}_{PB}}{d\tau} + \frac{d\boldsymbol{r}_{BA}}{d\tau} \quad (20)$$

$$\frac{d\boldsymbol{r}_{PB}}{d\tau} = \begin{bmatrix} \dfrac{dx_{PB}}{d\tau} \\ \dfrac{dy_{PB}}{d\tau} \end{bmatrix} = \dot{r}_{PB} \begin{bmatrix} \cos(\beta+f) \\ \sin(\beta+f) \end{bmatrix} - r_{PB} \begin{bmatrix} \sin(\beta+f) \\ -\cos(\beta+f) \end{bmatrix} \frac{df}{dt} \quad (21)$$

Based on the assumption of instantaneous disruption, $d\boldsymbol{r}_{BA}/d\tau$ is given by

$$\frac{d\boldsymbol{r}_{BA}}{d\tau} = \begin{bmatrix} \dfrac{dx_{BA}}{d\tau} \\ \dfrac{dy_{BA}}{d\tau} \end{bmatrix} = -r_{BA} \begin{bmatrix} \sin(\beta+\theta+f) \\ -\cos(\beta+\theta+f) \end{bmatrix} (\omega + \frac{df}{dt}) \quad (22)$$

Where $\omega$ is the angular velocity that asteroid A orbits around asteroid B before disruption.

Substitution of equations (21) and (22) into equation (20) gives

$$\begin{aligned} \frac{d\boldsymbol{r}_{PA}}{d\tau} &= \frac{d\boldsymbol{r}_{PB}}{d\tau} + \frac{d\boldsymbol{r}_{BA}}{d\tau} = \dot{r}_{PB} \begin{bmatrix} \cos(\beta+f) \\ \sin(\beta+f) \end{bmatrix} - r_{PB} \begin{bmatrix} \sin(\beta+f) \\ -\cos(\beta+f) \end{bmatrix} \frac{df}{dt} \\ &\quad - r_{BA} \begin{bmatrix} \sin(\beta+\theta+f) \\ -\cos(\beta+\theta+f) \end{bmatrix} (\omega + \frac{df}{dt}) \end{aligned} \quad (23)$$

Because the absolute velocity in the inertial frame is equal to its relative velocity plus the cross product of angular velocity of the moving system with the position vector. The velocity of asteroid *A* with respect to the planet can be described by the velocity in the rotating frame, which is given by

$$\frac{d\boldsymbol{r}_{PA}}{d\tau} = \frac{\tilde{d}\boldsymbol{r}_{PA}}{d\tau} + \omega_P \times \boldsymbol{r}_{PA} = \begin{bmatrix} \dot{x}_A - y_A \\ \dot{y}_A + x_A \end{bmatrix} \quad (24)$$

Where $\omega_P$ is the angular velocity of the planet around the sun.

From equations (19) and (24), we can solve the velocity in the rotating frame

$$\begin{pmatrix} \dot{x}_A \\ \dot{y}_A \end{pmatrix} = \dot{r}_{PB} \begin{bmatrix} \cos(\beta + f) \\ \sin(\beta + f) \end{bmatrix} + r_{PB} \begin{bmatrix} \sin(\beta + f) \\ -\cos(\beta + f) \end{bmatrix}(1 - \frac{df}{dt})$$
$$+ r_{BA} \begin{bmatrix} \sin(\beta + \theta + f) \\ -\cos(\beta + \theta + f) \end{bmatrix}(1 - \omega - \frac{df}{dt}) \quad (25)$$

Substitution of equations (18) and (25) into the expression of the energy integral gives

$$\dot{C}_A = \frac{\mu_B}{r_{BA}^2}(-\dot{r}_{PB}\cos\theta + r_{PB}\sin\theta(1 - \frac{df}{dt})) \quad (26)$$

The energy integral change of asteroid *A* during the flyby can be obtained by integrating equation (26)

$$\delta C = \frac{\mu_B}{r_{BA}^2} \int_{\tau_0}^{\tau_t}(-\dot{r}_{PB}\cos\theta + r_{PB}\sin\theta(1 - \frac{df}{dt}))d\tau \quad (27)$$

Based on assumption that the disruption occurs in the vicinity of the perigee, the integral region of equation (27) can be limited to the period of time when the binary passes the perigee.

$$\delta C = \int_{\tau_0}^{\tau_t}\dot{C}d\tau = \frac{\mu_B}{r_{BA}^2}\int_{\tau_f - \delta t}^{\tau_f + \delta t}(-\dot{r}_{PB}\cos\theta + r_{PB}\sin\theta(1 - \frac{df}{dt}))d\tau \quad (28)$$

where $\tau_f$ is the moment when the binary reaches the perigee. Because the disruption is instantaneous, asteroid B can be regarded as always being at the perigee. equation (28) can be simplified as

$$\delta C = \int_{\tau_0}^{\tau_t} \dot{C} d\tau = \frac{\mu_B}{r_{BA}^2} \int_{\tau_f-\delta t}^{\tau_f+\delta t} (-(\dot{r}_{PB})_{\tau_f} \cos\theta + (r_{PB})_{\tau_f} \sin\theta (1 - \left(\frac{df}{dt}\right)_{\tau_f})) d\tau \quad (29)$$

Because the motion of the binary around the planet is hyperbolic, the following relations hold:

$$\begin{cases} (r_{PB})_{\tau_f} = \left(\frac{p}{1+e\cos f}\right)_{\tau_f} = r_p \\ (\dot{r}_{PB})_{\tau_f} = \left(\frac{pe\sin f}{(1+e\cos f)^2} \frac{df}{d\tau}\right)_{\tau_f} = 0 \\ \left(\frac{df}{d\tau}\right)_{\tau_f} = \left(\frac{\sqrt{p\mu_P}}{r_{PB}^2}\right)_{\tau_f} = \frac{\sqrt{r_p(1+e)\mu_P}}{r_p^2} \end{cases} \quad (30)$$

Substitution of equation (30) into (28) gives

$$\delta C = \frac{\mu_B}{r_{BA}^2} \int_{\tau_f-\delta t}^{\tau_f+\delta t} ((r_p \sin\theta (1 - \frac{\sqrt{r_p(1+e)\mu_P}}{r_p^2})) d\tau \quad (31)$$

The independent variable can be changed from time to $\theta$. Thus, the integration can be rewritten as

$$\delta C = \frac{\mu_B}{r_{BA}^2} \left( \sqrt{\frac{(1+e)\mu_P}{r_p}} - r_p \right) \sqrt{\frac{r_{BA}^3}{\mu_B}} \cos(\theta_0 + \omega t)\big|_{\tau_f-\delta t}^{\tau_f+\delta t} \quad (32)$$

$$\delta C = \sqrt{\frac{\mu_B}{r_{BA}}} \left( \sqrt{\frac{(1+e)\mu_P}{r_p}} - r_p \right) \cos(\theta_t + \delta\theta) - \cos(\theta_t - \delta\theta)$$
$$= 2\sqrt{\frac{\mu_B}{r_{BA}}} \left( r_p - \sqrt{\frac{(1+e)\mu_P}{r_p}} \right) \sin(\theta_t)\sin(\delta\theta) \quad (33)$$

We have obtained an analytic expression to describe the energy exchange between asteroids. Because $(r_p - \sqrt{1+e)\mu_P / r_p}) < 0$ when the binary is in the SOI of the planet, the energy integral of asteroid $A$ increases when $\sin\theta_t < 0$ and decreases when $\sin\theta_t > 0$. $|\delta C|$ reaches the maximum when $\theta_t = \pm\pi/2$.

**4.2 Relative error between the analytic and numerical results**

To evaluate the validity of the analytic expression, the analytic results are compared to the numerical results. The relative error between the analytic and numerical results is defined as

$$C_{err} = \left|\frac{\delta C_a - \delta C_n}{\delta C_n}\right| \tag{34}$$

where $\delta C_n$ is the numerical result and $\delta C_a$ is the analytic result.

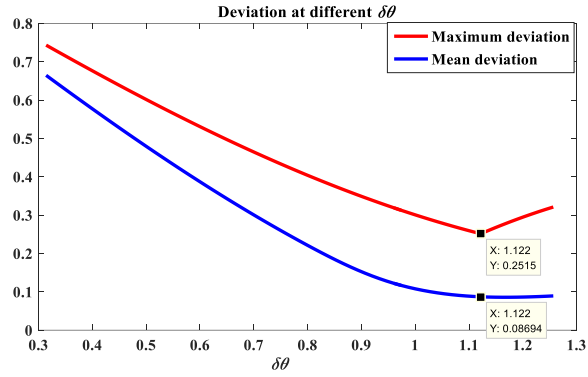

Figure 9. Deviation between the analytic and numerical results at different $\delta\theta$

As shown in the Figure 9, by calculating the analytic results at ergodic value of $\delta\theta(0 \sim 0.5\pi)$, we find that $C_{err}$ takes minimum value at $\delta\theta = 0.3571\pi$.

Figure 10 gives the analytic and numerical results for the case of $\theta_t = 0.7\pi$, $r_p = 0.6r_{td}$, $\mu_B = 3 \times 10^{-11}$, $\delta\theta = 0.3571\pi$. The maximum deviation between them is always less than 25% and the mean deviation is about 8.69%. Proving that analytical derivation of the previously is trustworthy.

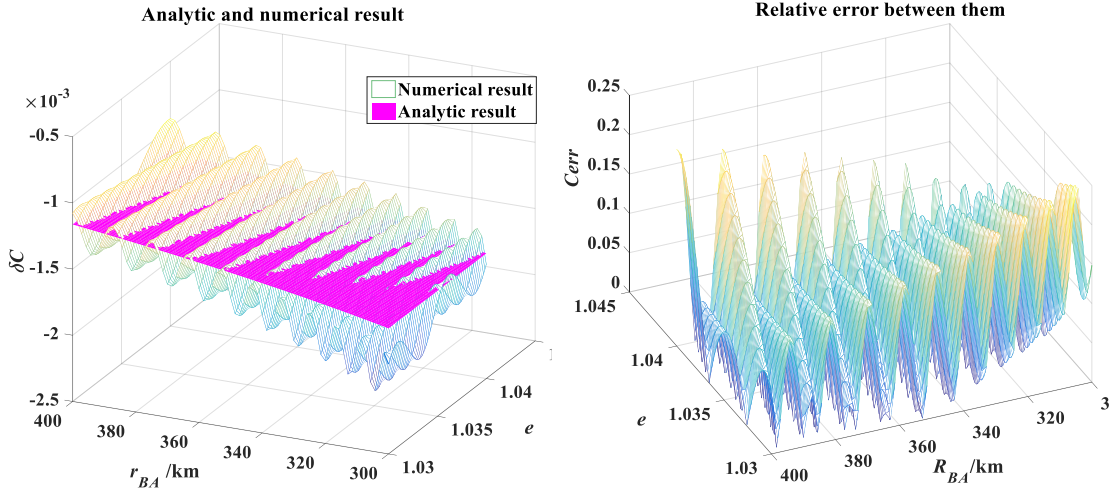

Figure 10. Relative error between the analytic and numerical results

## 5 Numerical simulation

Given the initial values when the binary enters the SOI of the planet, the energy integral change between the asteroids can be obtained by integrating equations (8) and (9) simultaneously.

We can obtain the true anomaly of the binary when it arrives at the SOI.

$$f_0 = -\arccos(r_p(1+e)/e/R_{SOI} - 1/e) \qquad (35)$$

where $R_{SOI}$ is the radius of the planet's SOI. Because asteroid $B$ moves around the planet in a hyperbolic orbit throughout the process and the distance between asteroids is invariant before the disruption, the initial phase $\theta_0$ between the asteroids can be expressed by $\theta_t$.

$$\begin{cases} \omega = \sqrt{\mu_B / r_{BA}^3} \\ T = 2\pi / \omega \\ t = \int_{f_0}^{0} \dfrac{1}{(1+e\cos f)^2} df \dfrac{r_p(1+e)^{3/2}}{\mu_P^{1/2}} \\ n = t/T \\ \varphi = (n - floor(n)) \times 2\pi \\ \theta_0 = \theta_t - \varphi \end{cases} \quad (36)$$

where $\omega$ is the angular velocity of asteroid A around B before the disruption, $t$ is the time that asteroid B arrives at the perigee, and $floor(n)$ is the nearest integer that is less than $n$. Thus, we can obtain the initial position and velocity of the asteroids relative to the planet.

$$\begin{cases} x_B = R_{SOI} \cos(\beta + f_0) \\ y_B = R_{SOI} \sin(\beta + f_0) \\ \dot{x}_B = \dfrac{r_p(1+e)e\sin(f_0)\sqrt{r_p(1+e)\mu_P}}{(1+e\cos(f_0))^2 R_{SOI}^2} \cos(\beta + f_0) + (1 - \dfrac{\sqrt{r_p(1+e)\mu_P}}{R_{SOI}^2}) R_{SOI} \sin(\beta + f_0) \\ \dot{y}_B = \dfrac{r_p(1+e)e\sin(f_0)\sqrt{r_p(1+e)\mu_P}}{(1+e\cos(f_0))^2 R_{SOI}^2} \sin(\beta + f_0) - (1 - \dfrac{\sqrt{r_p(1+e)\mu_P}}{R_{SOI}^2}) R_{SOI} \cos(\beta + f_0) \end{cases}$$

(37)

$$\begin{cases} x_A = R_{SOI} \cos(\beta + f_0) + r_{BA} \cos(\beta + f_0 + \theta_0) \\ y_A = R_{SOI} \sin(\beta + f_0) + r_{BA} \sin(\beta + f_0 + \theta_0) \\ \dot{x}_A = \dfrac{r_p(1+e)e\sin(f_0)\sqrt{r_p(1+e)\mu_P}}{(1+e\cos(f_0))^2 R_{SOI}^2} \cos(\beta + f_0) + (1 - \dfrac{\sqrt{r_p(1+e)\mu_P}}{R_{SOI}^2}) R_{SOI} \sin(\beta + f_0) \\ \quad + r_{BA} \sin(\beta + f_0 + \theta_0)(1 - \omega - \dfrac{\sqrt{r_p(1+e)\mu_P}}{R_{SOI}^2}) \\ \dot{y}_A = \dfrac{r_p(1+e)e\sin(f_0)\sqrt{r_p(1+e)\mu_P}}{(1+e\cos(f_0))^2 R_{SOI}^2} \sin(\beta + f_0) - (1 - \dfrac{\sqrt{r_p(1+e)\mu_P}}{R_{SOI}^2}) R_{SOI} \cos(\beta + f_0) \\ \quad - r_{BA} \cos(\beta + f_0 + \theta_0)(1 - \omega - \dfrac{\sqrt{r_p(1+e)\mu_P}}{R_{SOI}^2}) \end{cases}$$

(38)

So far, given the values of $\beta$, $\theta_t$, $\mu_B$, $r_p$, $e$ and $r_{BA}$, the initial values we need to integrate equations (8) and (9) can be obtained.

**5.1 Cases for capture/escape scenarios**

Considering two scenarios, $\theta_t = 0.7\pi, \mu_B = 3\times 10^{-11}, r_{BA} = 2.33333\times 10^{-6}\ \text{AU}, e = 1.04$, $r_p = 0.7 r_{td}$ and $\theta_t = 0.8\pi, \mu_B = 3\times 10^{-11}, r_{BA} = 2.1333\times 10^{-6}\ \text{AU}, e = 1.02$, $r_p = 0.6 r_{td}$. As shown in Figure 11 and Figure 12, the flyby reduces the energy integral of asteroid $A$ below the critical value and makes its orbit bounded around the planet. In these two scenarios, asteroids $A$ is captured as a result of energy exchange.

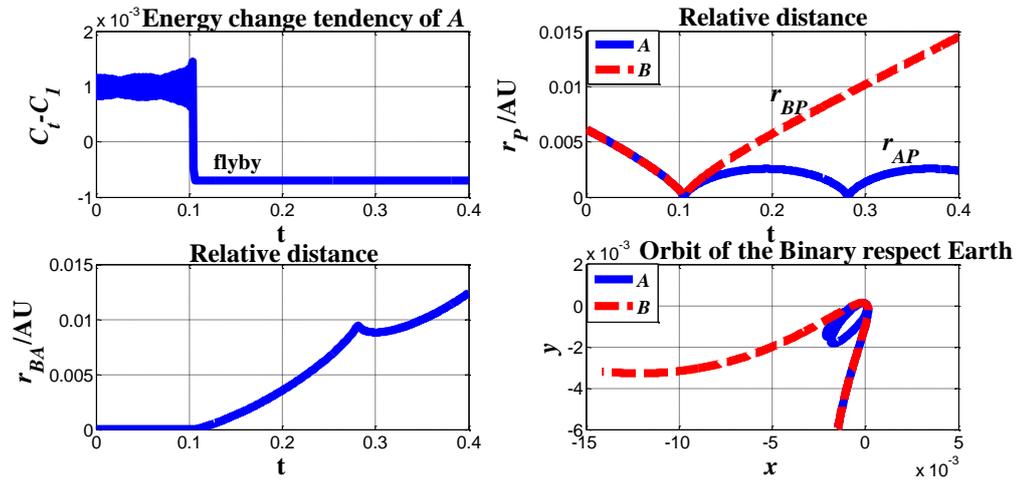

Figure 11. Capture scene $\theta_t = 0.7\pi$, $r_p = 0.7 r_{td}$, $e = 1.04$, $r_{BA} = 2.3333\times 10^{-6}$

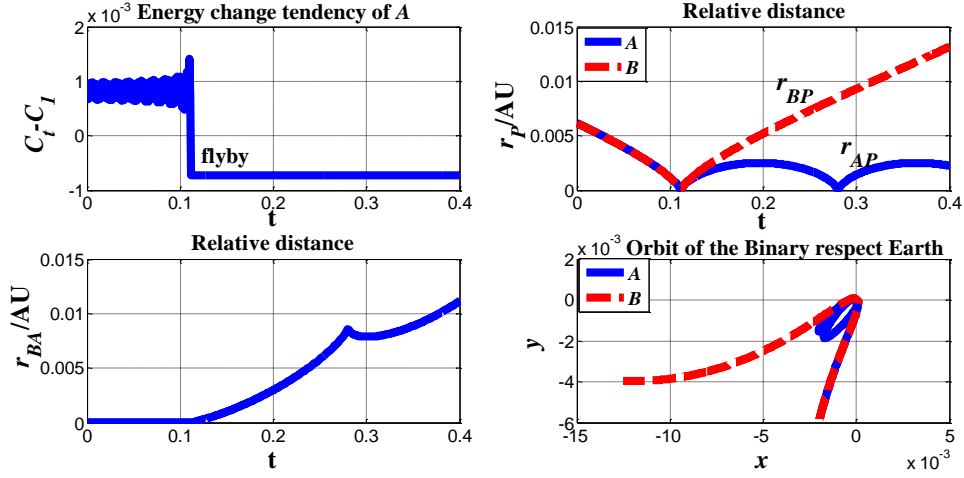

Figure 12. Capture scene $\theta_t = 0.8\pi$, $r_p = 0.6r_{td}$, $e = 1.02$, $r_{BA} = 2.1333 \times 10^{-6}$

Similarly, consider other two scenarios, $\theta_t = 0.7\pi$, $\mu_B = 3 \times 10^{-11}$, $r_p = 0.6r_{td}$, $e = 1.06$, $r_{BA} = 2.6667 \times 10^{-6}$ AU and $\theta_t = 0.7\pi$, $\mu_B = 3 \times 10^{-11}$, $r_p = 0.7r_{td}$, $e = 1.07$, $r_{BA} = 2.6667 \times 10^{-6}$ AU. As shown in Figure 13 and Figure 14, the energy integral of asteroid $A$ is above the critical value after flyby. However, the separation between the binary is above the Hill radius of asteroid $B$. Thus, the binary is disrupted but not captured in these two scenarios.

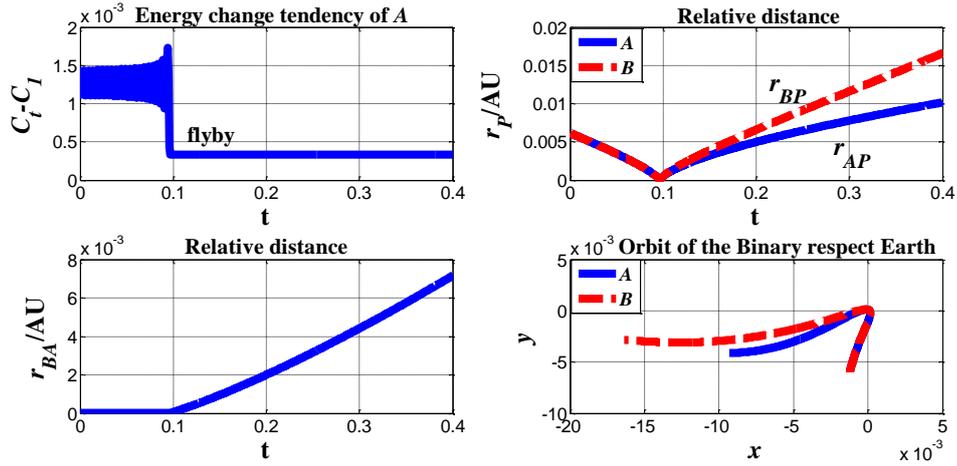

Figure 13. Escape scene $\theta_t = 0.7\pi$, $r_p = 0.6r_{td}$, $e = 1.06$, $r_{BA} = 2.6667 \times 10^{-6}$

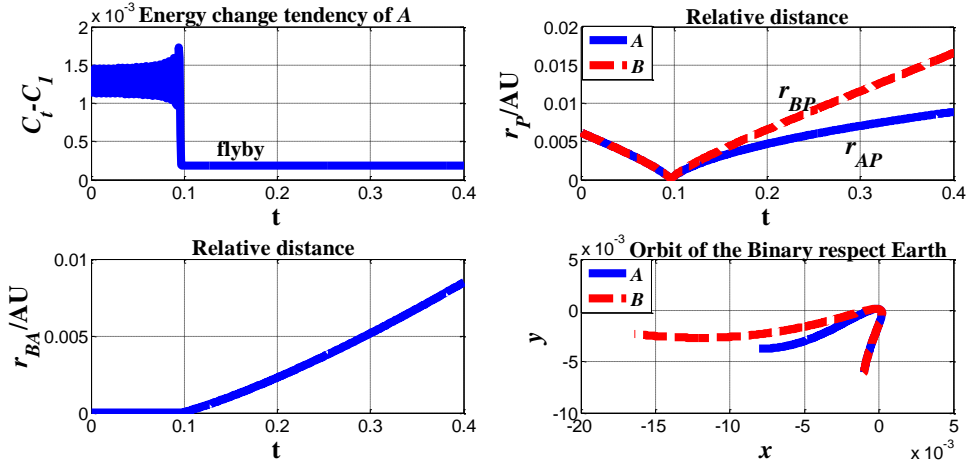

Figure 14. Escape scene $\theta_t = 0.7\pi$, $r_p = 0.7 r_{td}$, $e = 1.07$, $r_{BA} = 2.6667 \times 10^{-6}$

## 5.2 Capture region

In this section, the capture regions of asteroid *A* are studied for different pairs of parameters to illustrate the influences of each parameter on these regions.

Given $\beta = 0.3\pi$, $\mu_B = 3 \times 10^{-11}$, $r_p = 0.6 r_{td}$, we can obtain the capture regions represented in the $e - r_{BA}$ space for different values of $\theta_t$. As shown in Figure 15, the region of dashed line is the capture region where the energy integral of asteroid *A* is below the critical value.

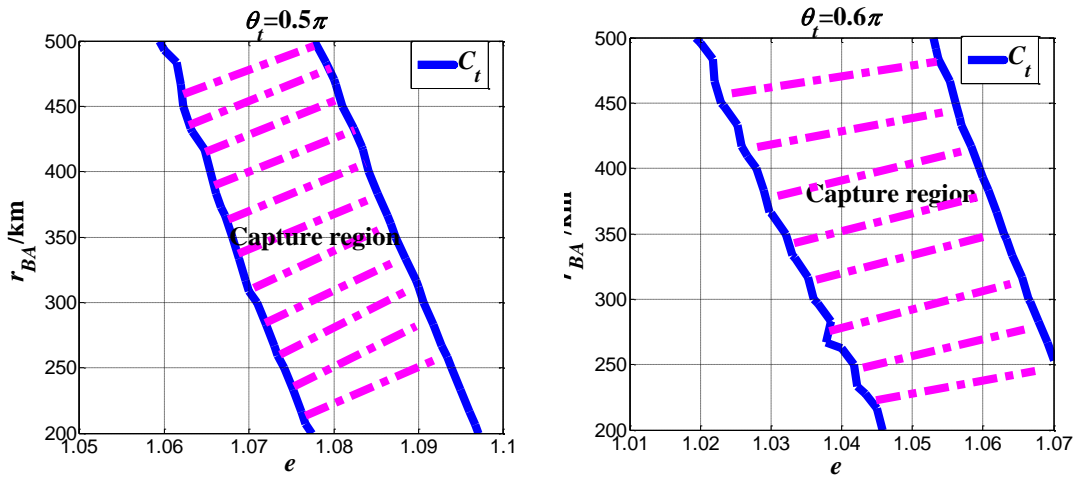

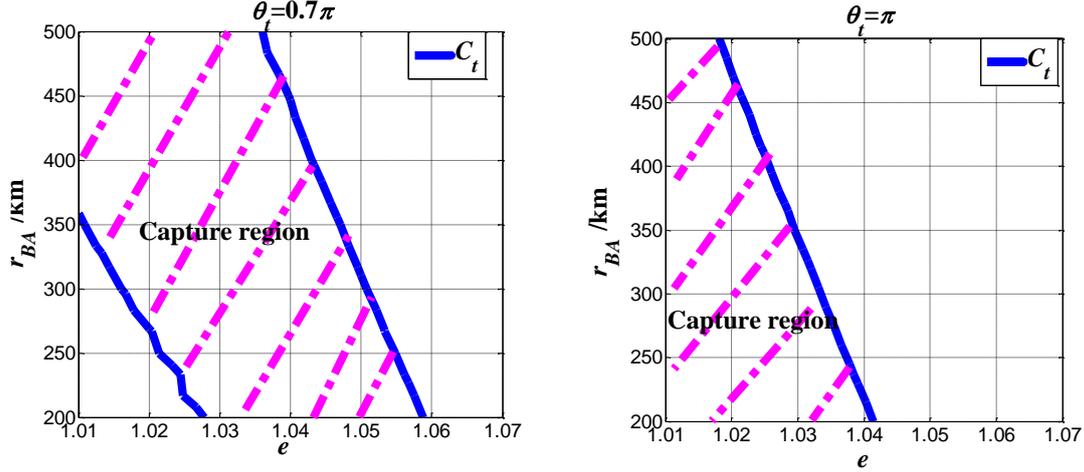

Figure 15. The capture region for different $\theta_t$

From the analytic expression in equation (33), the absolute value of energy integral variation reaches the maximum when $\theta_t = \pi/2$. However, the capture region we have obtained expands as $\theta_t$ approaches $0.7\pi$, which is inconsistent with the analytic conclusion. We assume that the binary is disrupted instantaneously and the distance between the asteroids is invariant before reaching the perigee. However, $r_{BA}$ is time-varying actually, which leads to hypothetical phase angle $\theta_t$ between the asteroids at the perigee deviating from its actual phase.

From the analytic expression, the parameters that influence the capture region include $\theta_t$, $e$, $r_{BA}$, $\mu_B$, $r_p$ and $\mu_P$.

1) Influence of $\mu_B$ (mass of asteroid $B$)

The derivative of $\mu_B$ can be obtained using equation (33). Formula $\dfrac{\partial \delta C}{\partial \mu_B} < 0$ always works when $\sin(\theta_t) > 0$. Thus, the energy integral change $|\delta C|$ increases with $\mu_B$. The capture region should expand as $\mu_B$ increases.

Given $\theta_t = 0.5\pi$, $r_p = 0.6 r_{td}$, the capture regions are obtained for $\mu_B = 2 \times 10^{-11}$,

$3\times10^{-11}$, $4\times10^{-11}$, and $5\times10^{-11}$, respectively. As shown in Figure 16, the capture region expands as expected.

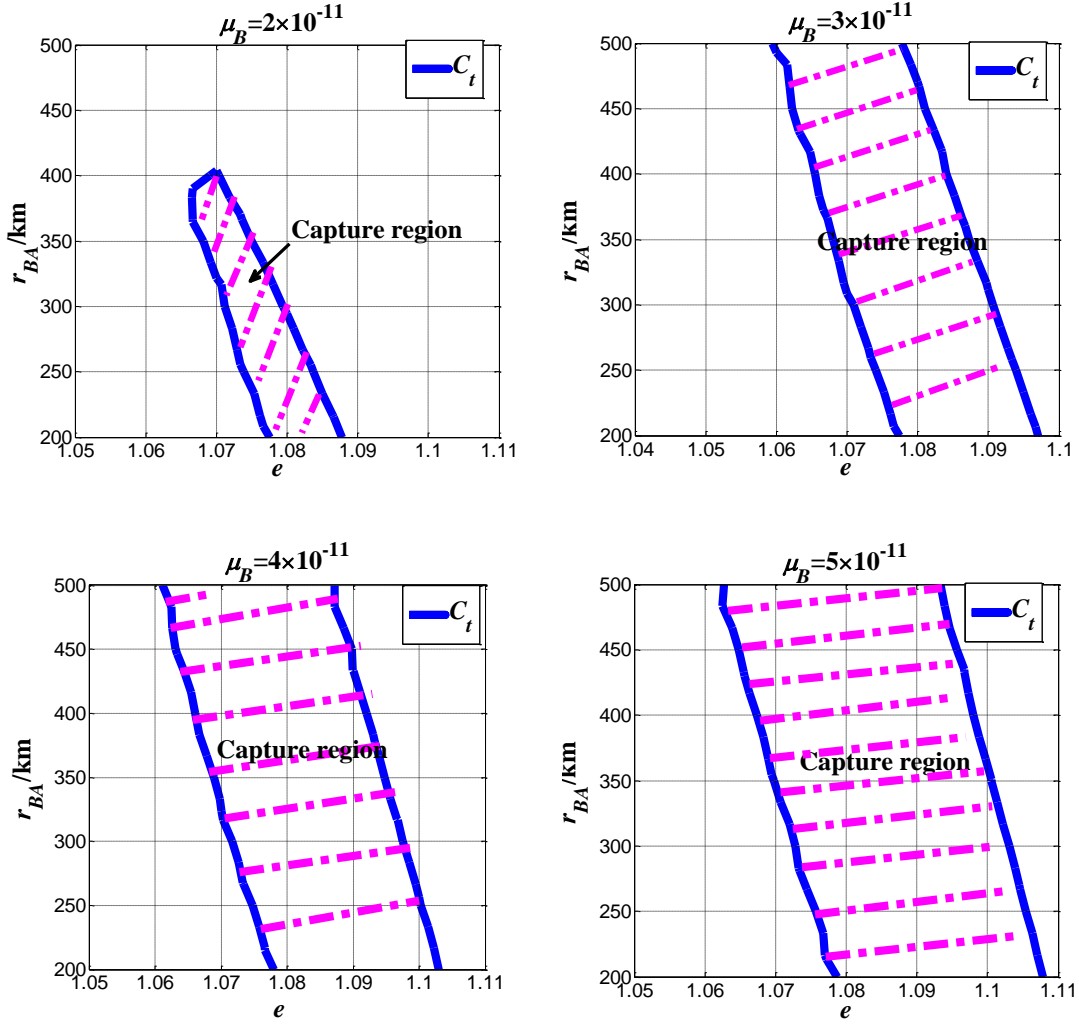

Figure 16. Capture region at different $\mu_B$

2) Influence of $r_p$ (perigee radius)

The derivative of $r_p$ is obtained using the equation (33). The following formula always works as $\sin(\theta_t) > 0$.

$$\frac{\partial \delta C}{\partial r_p} = 2\sqrt{\frac{\mu_B}{r_{BA}}}\sin(\theta_t)\sin(\delta\theta)\left(1+\frac{1}{2}\sqrt{\frac{(1+e)\mu_P}{r_p^3}}\right) > 0$$

Given $\delta C < 0$ when $\sin(\theta_t) > 0$, the energy integral change $|\delta C|$ decreases with $r_p$. Thus, the capture region should shrinks as $r_p$ increases.

Given $\mu_B = 3 \times 10^{-11}$, $\theta_t = 0.7\pi$, the capture regions are obtained for $r_p = 0.7 r_{td}$, $0.8 r_{td}$, respectively. As shown in Figure 17, the capture region shrinks as expected.

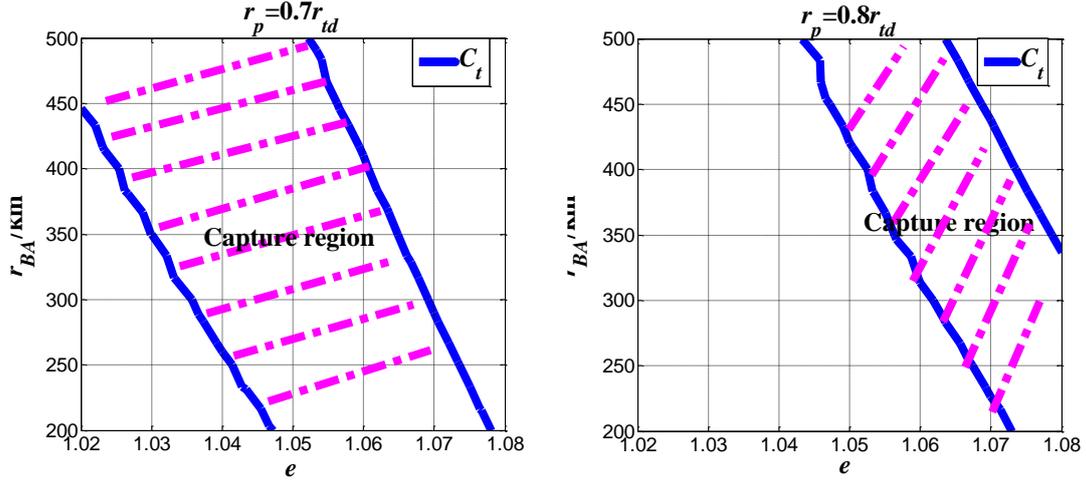

Figure 17. Capture region at different $r_p$

3) Influence of $\mu_P$ (mass of planet)

From equation (33) and the expression of energy integral $C_1$, we know that $C_1$ and $\delta C$ both decrease as $\mu_P$ increases. The change rates of $C_1$ and $C_t$ are determined numerically. As shown Figure 18, the value of $C_t - C_1$ increases with $\mu_P$. Thus, the capture region will shrink as $\mu_P$ increases.

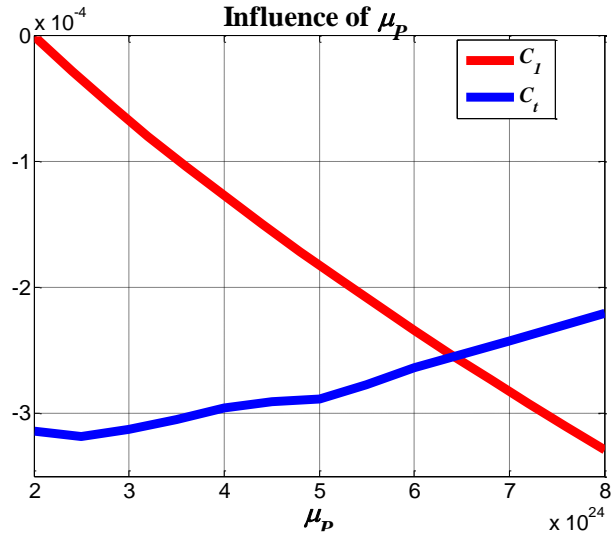

Figure 18. The change tendency of $C_l$ and $C_t$ with the increase of $\mu_P$

Given $\theta_t = 0.7\pi$, $r_p = 0.7 r_{td}$, the capture regions for different values of $\mu_P$ are obtained. As shown in Figure 19, the capture region shrinks as $\mu_P$ increases.

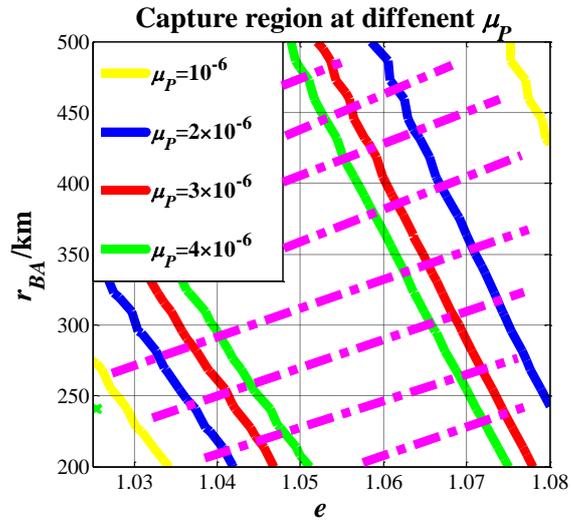

Figure 19. Capture region at different $\mu_P$

# 6 Conclusions

The disruption/capture mechanism of the sun-planet-binary system is investigated. We obtain the disruption region of the binary and draw two important conclusions by numerical simulations: (1) the binary is disrupted instantaneously in the vicinity of the perigee; (2) the binary' original orbit almost keeps undisturbed before the disruption. For the capture problem, an analytic expression to describe the energy exchange between the asteroids is derived. We can describe the energy exchange between the binary intuitively by this analytic expression intuitively. To verify the validity of the analytic expression, the analytic and numerical results of the exchange energy integral are obtained, and the maximum deviation between them is always less than 25% and the mean deviation is about 8.69%. Several numerical examples are given to illustrate the capture and escape processes. We obtain the capture region in the $e - r_{BA}$ space and analyse the influences of $r_p$, $\mu_P$ and $\mu_B$ on the capture region.

## Acknowledgements

The authors would like to acknowledge the support from the National Natural Science Foundation of China (Grants No. 11272004) and the National Basic Research Program of China (973 Program, 2012CB720000).